\documentclass[aps,preprint]{revtex4}

\usepackage{amsmath,amssymb,bm}

\def\({\left (}
\def\){\right )}
\def\[{\left[}
\def\]{\right]}
\def\mcA{\mathcal{A}}
\def\mcF{\mathcal{F}}
\def\Mk{M_{KK}}
\def\tB{\tilde{B}}
\def\hA{\hat{A}}
\def\hF{\hat{F}}
\def\half{\frac{1}{2}}
\def\tr{\textrm{tr}}
\def\Ai{A^{(i)}}
\def\Fi{F^{(i)}}
\def\Aj{A^{(j)}}

\def\Ak{A^{(k)}}

\def\Fimn{F^{(i)\mu \nu}}
\def\Fimz{F^{(i)\mu z}}
\newcommand\x{{\mathbf{x}}}
\renewcommand\d{\partial}

\begin{document}

\preprint{INT PUB 08-10}

\title{Magnetized baryonic matter in holographic QCD}
\author{Ethan G.~Thompson} 
\affiliation{Department of Physics, University of Washington, Seattle, 
Washington, WA 98195-1560, USA}
\author{Dam T.~Son}
\affiliation{Institute for Nuclear Theory, University of Washington,
Seattle, Washington 98195-1550, USA}
\date{June 2008}

\begin{abstract}

We investigate the properties of the Sakai-Sugimoto model at finite
magnetic field and baryon chemical potentials.  We show that in a finite
magnetic field, there exists a spatially homogeneous configuration
carrying finite baryon number density.  At low magnetic field and
baryon chemical potential the equation of state of the matter
coincides with that obtained from the chiral perturbation theory
Lagrangian with an anomalous term.  We discuss the behavior of the
system at larger magnetic fields.

\end{abstract}

\maketitle


\section{Introduction}

Recently, gauge/gravity 
duality~\cite{Maldacena:1997re,Gubser:1998bc,Witten:1998qj} 
has been used extensively to
investigate properties of strongly coupled gauge theories at finite
temperature and/or density.  The popularity of this method is due to its
ability to calculate in the strong coupling regime.  The main disadvantage
is that the models that can be solved, for example the ${\cal N}=4$
super Yang-Mills (SYM) theory, do not coincide with QCD, typically
containing additional degrees of freedoms like adjoint fermions or
scalars.

While finite-temperature ${\cal N}=4$ SYM plasma has many features
reminiscent of QCD plasma at temperatures not
too large compared to the deconfinement temperature, it is more
difficult to construct a holographic model of cold nuclear or quark
matter.  One problem is that at large $N_c$ nuclear matter is a
crystal instead of a liquid.  This fact finds reflection in the
Sakai-Sugimoto model~\cite{Sakai:2004cn}, where baryons are 5D
instanton particles and nuclear matter is a crystal of such
instantons~\cite{Kim:2007zm}, which is necessarily 
inhomogeneous~\cite{Rozali:2007rx}.

In this paper, we investigate the possible gravity dual of magnetized
nuclear matter.  In a recent study~\cite{Son:2007ny}, it was found
that in the chiral limit of massless quarks, at any magnetic field the
ground state of finite-density matter is not a crystal, but a
spatially homogeneous phase.  At low density such a phase is
characterized by a finite gradient of the $\pi^0$ field,
$\nabla\pi^0\neq0$ [in the parametrization where the chiral condensate
is proportional to $\exp(i\pi^0\tau^3/f_\pi)$].  At finite quark
masses this state becomes a stack of $\pi^0$ domain walls.  The
$\pi^0$ domain wall is locally stable in magnetic fields stronger than
$B_0$, and is energetically more favorable than nuclear matter above a
magnetic field $B_1$, where both $B_0$ and $B_1$ vanish in the chiral
limit.

The treatment of Ref.~\cite{Son:2007ny} relies on the use of chiral
perturbation theory, including the appropriate WZW term in the
presence of an electromagnetic field and a baryon number chemical
potential.  It is valid only for sufficiently small magnetic field
and baryon chemical potential.  

In this paper, we search for a similar solution in the Sakai-Sugimoto
model of holographic QCD~\cite{Sakai:2004cn}.  The Sakai-Sugimoto
model is an application of the AdS/CFT conjecture involving a system
of $N_{\!f}$ D8--$\overline{\textrm{D8}}$ probe brane pairs in a D4
brane background of type IIA string theory.  The model exhibits chiral
symmetry breaking and thereby reproduces much of the low-energy
physics of massless QCD, such as the octet of pseudoscalar
Nambu-Goldstone bosons.  Including the Chern-Simons (CS) term of the
probe brane action is equivalent to including the effects of the axial
anomaly on the field theory side.  While the model has been used to
investigate properties of the vacuum and the thermal state (with zero
chemical potentials) in an external magnetic
field~\cite{Bergman:2008sg,Johnson:2008vn}, the case when both the
magnetic field and the baryon chemical potentials are nonzero has not
been considered.

Because the Sakai-Sugimoto model incorporates the axial anomaly into a
theory of massless pions, one would expect to find, at least at small
magnetic fields and baryon chemical potentials, a solution similar to
the one found in~\cite{Son:2007ny}.
The purpose of this paper is to demonstrate that
solution.

We found, as expected, that at low $B$ the results of
Ref.~\cite{Son:2007ny} are reproduced.  At larger values of the
magnetic field, the quadratic approximation to the DBI action of the
probe branes can no longer be trusted.  However, we can still consider
the quadratic action at large $B$ as a bottom-up AdS/QCD
theory~\cite{Son:2003et,Erlich:2005qh,Da Rold:2005zs}.  In this sense, we
unexpectedly discovered that in the opposite limit of large $B$ the
zero-temperature thermodynamics of matter with finite baryon density
is identical to the thermodyanamics of free quarks, which fill energy
levels in the lowest Landau level.  It is rather surprising given that
the supergravity limit corresponds to the strong coupling regime in
field theory.

In Sec.~\ref{sec:SS}, the necessary facets of the Sakai-Sugimoto model
are reviewed.  In Sec.~\ref{sec:solution}, the domain wall solution is
presented.  We give interpretations of the result in
Sec.~\ref{sec:interpretation}.  Section~\ref{sec:conclusion} provides
a concluding discussion.

{\sl Note added:} While this paper was being completed, we learned
that a similar investigation was also being carried out
in~\cite{Bergman}, which contains some overlapping results.

\section{Review of the Sakai-Sugimoto model}
\label{sec:SS}

In~\cite{Witten:1998zw}, a way to holographically model
nonsupersymmetric pure Yang-Mills theory was presented.  Although the
initial motivation of the model involved the M-theory duality on
AdS$_7\times$S$^4$, Witten argued that it could equivalently be
described as the background of a stack of $N_c$ D4 branes in type IIA
supergravity, where one of the directions parallel to the D4 branes is
compactified into a circle.  Antiperiodic boundary conditions around
the circle are imposed on the fermionic fields, giving the fermions a
mass and breaking the supersymmetry.  The scalar fields also acquire a
mass at the one-loop level, leaving the $SU(N_c)$ vector field as the
only massless field in the theory, and thus reproducing
nonsupersymmetric pure Yang-Mills theory at energies small compared 
to the Kaluza-Klein scale $\Mk$.


Sakai and Sugimoto added massless flavor to the theory by considering
the addition of $N_{\!f}$ D8--$\overline{\textrm{D8}}$ probe branes to
the background~\cite{Sakai:2004cn}, where the probe branes are 
transverse to the circle of compactification.  The essential idea of adding
flavor to holographic systems via probe branes~\cite{Karch:2002sh} is
that if $N_{\!f} \ll N_c$, the backreaction of the probe branes on the
geometry can be neglected, and the probe brane action consists simply
of the Dirac-Born-Infeld (DBI) action in the original background, plus
the relevant Chern-Simons (CS) terms.

There is a U($N_{\!f}$)$\times$U($N_{\!f}$) gauge symmetry living on
the probe brane pairs, which provides a global chiral symmetry on
the field theory side.  The geometry of the D4 branes is cigar-shaped,
and one finds from analysis of the DBI action that the
D8--$\overline{\textrm{D8}}$ branes merge at some value of the radial
coordinate.  Thus, the U($N_{\!f}$)$\times$U($N_{\!f}$) symmetry is
broken to a single U($N_{\!f}$); this is a holographic manifestation
of chiral symmetry breaking.  It was shown in
Refs.~\cite{Sakai:2004cn,Sakai:2005yt} that the DBI action of the
probe branes reproduces much of the low-energy physics of QCD,
including the chiral Lagrangian and qualitative features of the meson
spectrum.  Inclusion of the CS term for the probe branes incorporates
the effects of the axial anomaly into the low energy theory.

It is possible to construct baryons in this model~\cite{Hata:2007mb}.
Witten's baryon vertex appears as a D4 brane wrapping the
S$^4$.  Strings stretching between the D4 brane and the D8 branes will 
source the gauge field living on the D8s.  The baryon number is then 
given in terms of the SU($N_{\!f}$)
valued field strengths $F$ living on the probe branes as
\begin{equation}\label{baryonnumber}
  N_B = \frac{1}{32 \pi^2} \int d^3x dz ~ \epsilon_{MNPQ} \tr 
  \[ F_{MN} F_{PQ} \],
\end{equation}
where $M, N, P, Q = 1, 2 ,3, z$.  The baryon number integral also shows up in the
probe brane action in the CS term coupled to the time-component of the
$U(1)$ part of the gauge field, $\hA$:
\begin{equation}
S_{CS} \supset \frac{N_c}{64 \pi^2} \epsilon_{MNPQ} \int d^4x dz ~ \hA_0 \tr 
  \[F_{MN} F_{PQ} \].
\end{equation}  
Thus, $\hA_0$ acts as a source for baryon number, and, in
order to turn on a finite chemical potential, we will consider
solutions with nontrivial $\hA_0$.

We also wish to turn on an external magnetic field that couples to our flavor degrees of freedom.  
There is no proper U(1) gauge field in our theory.  However, we can simulate the effects of an external field by weakly gauging U(1) subgroups of the global chiral symmetry.  
In real QCD with $N_{\!f} = 2$ flavors, the electric charge is related
to both the third component of isospin and to the baryon number, $Q =
I_3 + \half B$.  Therefore, in order to properly introduce an
electromagnetic field into Sakai-Sugimoto, one would have to gauge
both one of the components of $I_3$ and the baryon number potential.
However, for simplicity, in this paper we will only consider gauging
of the isospin component by looking for solutions with a nonzero
$F^{(3)}$.

\section{Solution with finite magnetic field and chemical potential}
\label{sec:solution}

We will first establish our conventions and notation.  

In this paper, we will consider the case of two flavor D8 branes.  The branes and the antibranes will be maximally separated around the circle of compactification, as in the original treatment by Sakai and Sugimoto~\cite{Sakai:2004cn}.  

The effective action of the probe branes is written in
terms of a U(2) $=$ U(1)$_B \times$ SU(2) five-dimensional gauge
field. It will often be convenient to distinguish the fifth spacetime index
$z$ from the four-dimensional boundary indices.  Thus, upper-case
roman letters $M, N...$ run over all five spacetime directions,
whereas lower-case greek letters $\mu ,\nu...$ run only over $0, 1, 2,
3$.  We will use a combination of form and component notation to
describe the gauge fields.  The U(2)-valued form fields are given by
\begin{align}
  \mathcal{A} & =  \mathcal{A}_{M} dx^{M}, \\
  \mcF & =  d\mcA + i\mcA \wedge \mcA.
\end{align}
The gauge fields can be decomposed into the $U(1)$ part $\hA$ and the
$SU(2)$ part $A$ as
\begin{equation}
\mcA = A +  \frac{\hA}{2} = \Ai T^i + \frac{\hA}{2} \textbf{1},
\end{equation}
where the $T^i$ are the $SU(2)$ generators normalized as $\tr \[
T^iT^j \] = \half \delta^{ij}$ and the numerical factor on the $U(1)$
piece is to ensure that the $U(1)$ generator is normalized in the same
manner.  Likewise, $\mcF$ is decomposed in terms of $F$ and $\hF$:
\begin{equation}
\mcF = \Fi T^i + \frac{\hF}{2}.
\end{equation} 
In component form, 
\begin{align}
\Fi_{MN} & =  \partial_{M}\Ai_N - \partial_N \Ai_M 
  - \epsilon^{ijk}\Aj_M \Ak_N, \label{components}\\
\hF_{MN} & =  \partial_M \hA_N - \partial_N \hA_M.
\end{align}

The effective theory of the branes is described by the 
action~\cite{Sakai:2004cn}
\begin{subequations}\label{action}
\begin{align}
S & =  S_{YM} + S_{CS}, \\
\label{YM}S_{YM} & =  - \kappa\int\! d^4x\, dz\, 
  \tr \[ \frac{1}{2}h(z) \mcF_{\mu  \nu} 
  \mcF^{\mu \nu} + \Mk^2 k(z) \mcF_{\mu z} \mcF^{\mu z} \] 
  + \mathcal{O}(\mcF^3), \\
S_{CS} & =  \frac{N_c}{24 \pi^2} \int\! \tr \[ \mcA \mcF^2 - \frac{i}{2} 
  \mcA^3 \mcF - \frac{1}{10} \mcA^5 \],
\end{align}
\end{subequations}
where 
\begin{equation}
  \kappa = \frac{\lambda N_c}{216 \pi^3}\,, 
\end{equation}
and $h(z)$ and $k(z)$
are defined as
\begin{equation}
h(z) = (1+z^2)^{-1/3},\qquad k(z) = 1 + z^2.
\end{equation}
In ``bottom-up'' AdS/QCD models where the chiral symmetry is
spontaneously broken by the boundary conditions at the IR brane, the
action has the same form as~(\ref{action}), but with different
functions $h(z)$ and $k(z)$.  For example, in the model considered
in~\cite{Son:2003et}, $h(z)=\textrm{const}$ and $g(z)\sim \cosh(bz)$,
where $b$ is a constant. 

The Yang-Mills action (\ref{YM}) arises from expanding the DBI action 
for the probe branes to second order in field strengths.  For large values 
of the field strength, this expansion will no be longer valid.  
We will see later that turning on a magnetic involves setting 
$F^{(3)}_{12} = B$.  It can be shown that the cubic terms involving 
$B$ can be dropped as long as $B$ satisfies the inequality
\begin{equation}
\frac{27 \pi}{2 \lambda} \frac{B}{\Mk^2} \ll 1.
\end{equation}
Alternatively, the action (\ref{YM}) can be interpreted as a bottom-up 
effective action.  From this perspective, $B$ is allowed to be 
arbitrarily large.  

In terms of the $U(1)$ and $SU(2)$ pieces, the action reads~\cite{Hata:2007mb}
\begin{eqnarray}
 S_{YM}  &=&  -\frac{\kappa}{2} \int\!d^4x\, dz  \left[ \frac{h(z)}{2} 
 \( \Fi_{\mu \nu} \Fimn {+} \hF_{\mu \nu} \hF^{\mu \nu} \) + \Mk^2 k(z) 
 \(\Fi_{\mu z} \Fimz {+} \hF_{\mu z} \hF^{\mu z} \) \right] \\
 S_{CS}  &=&  \frac{N_c}{24 \pi^2} \int \( \frac{3}{2} \hA ~\tr F^2 
 + \frac{1}{4}\hA \hF^2 + \half d\(\hA ~\tr \[2FA - \frac{i}{2}A^3\]\) \).
\end{eqnarray}
In~\cite{Hata:2007mb}, a localized soliton solution is found whose size is
$\mathcal{O}(1/\sqrt{\lambda})$.  In this case, the scaling of the gauge field
solution allows the equations of motion to be expanded as a series in
$\lambda$.  In our case, there is no such scaling in our solution, so
we must work with the full equations of motion.  Variation of the
above action leads to the equations
\begin{align}
\frac8\alpha \left[ h(z) \partial_M \hat{F}^{MN} +
  \Mk^2\partial_z \bigl( k(z)\hat{F}^{zN}\bigr)\right] & = 
 -\epsilon^{N\nu \rho \sigma \lambda}( F^{(i)}_{\nu \rho} 
 F^{(i)}_{\sigma \lambda} +\hat{F}_{\nu \rho} \hat{F}_{\sigma \lambda}),\\
  \frac8\alpha \Mk^2 k(z) \partial_M \hat{F}^{Mz} & = 
 -\epsilon^{z \nu \rho \sigma \lambda}( F^{(i)}_{\nu \rho} 
 F^{(i)}_{\sigma \lambda} + \hat{F}_{\nu \rho} \hat{F}_{\sigma \lambda}),\\
   \frac4\alpha \left[ h(z) D_M F^{(i)MN} 
   + \Mk^2 \d_z \bigl( k(z)F^{(i)zN} \bigr) \right] 
  & = -\epsilon^{N \nu \rho \sigma \lambda} 
  \hat{F}_{\nu \rho} F_{\sigma \lambda}^{(i)},\\
  \frac4\alpha \Mk^2 k(z) D_M F^{(i)Mz}  & =  
 -\epsilon^{z \nu \rho \sigma \lambda} \hat{F}_{\nu \rho} 
 F_{\sigma \lambda}^{(i)},
\end{align}
where the covariant derivative acting on field strengths is 
\begin{equation}
D_\mu F^{(i)\mu \nu} = \partial_\mu F^{(i)\mu \nu} 
 +  \epsilon^{ijk}A_\mu^{(j)} F^{(k)\mu \nu}.
\end{equation}
and we have defined, for future convenience,
\begin{equation}
 \alpha = \frac{N_c}{16\pi^2\kappa} = \frac{27\pi}{2\lambda}\,.
\end{equation}

We wish to turn on a magnetic field and a baryon number chemical
potential and to look for a solution homogeneous in Minkowski space.
As discussed above, in order to turn on the magentic field we will assume a nonzero
$F^{(3)}_{12}$, and to generate a baryon number chemical potential we
will assume a nonzero $\hat{A}_0$. Recall that in terms of the five-dimensional
gauge fields, the baryon number is given by (\ref{baryonnumber}),
\begin{equation}
  N_B = \frac{1}{64\pi^2} \int\! d^3x\,dz\, 
  \epsilon^{MNPQ}F^{(i)}_{MN}F^{(i)}_{PQ}
\end{equation}
where $M, N, P, Q$ are not zero.  We see that in order to have nonzero
baryon number, if $F^{(3)}_{12}$ is nonzero then $F^{(3)}_{3z}$ must
also be nonzero.  Because we want a solution that is homogeneous in
the four-dimensional coordinates, all these quantities will only depend on $z$.  Let
us assume all other fields are zero.  To reduce clutter, we will
henceforth denote $F^{(3)}$ simply by $F$.

The second and fourth equations of motion are then trivially satisfied, 
whereas the first and third become
\begin{align}
 \frac{\Mk^2}\alpha \partial_z \bigl(k(z)\hat{F}^{z0}\bigr) &=  -F^{12}F^{3z}\\
 \frac{\Mk^2}\alpha (\partial_z \bigl(k(z){F}^{3z}\bigr) &= 
 -\hat{F}^{z0}F^{12}.
\end{align}
The covariant derivative has reduced to a partial derivative because 
$A^{(j)}$ is zero for $j= 1 ,2$.  


We notice that due to the Bianchi identity and the requirement that
fields depend only on $z$, $F_{12}$ has to be a constant, $F_{12} =
B$.  The equations can be solved exactly for any function $k(z)$.  For
$k(z)=1+z^2$ as in the Sakai-Sugimoto model, the general solution is
%
%
\begin{align} 
   F^{3z} &= c_1 \frac{\exp(\tB \arctan z)}{1+z^2} 
+ c_2 \frac{\exp(-\tB \arctan z)}{1+z^2},\label{F3z}\\
\hat{F}^{z0} & = -c_1 \frac{\exp(\tB \arctan z)}{1+z^2} 
+ c_2 \frac{\exp(-\tB \arctan z)}{1+z^2}\,. \label{Fz0} 
\end{align}
where $\tB = \alpha B/\Mk^2$.

In the presence of a finite chemical potential, the thermodynamic 
ground state will minimize the free energy  $H - \mu N_B$.  
Minimizing this quantity will give us the values of $c_1$ and $c_2$ 
in the ground state.  Under our Ansatz, the baryon 
number (\ref{baryonnumber}) reduces to
\begin{equation}
  N_B = \frac{1}{8 \pi^2} \int\! d^3x\,dz\,  B F^{3z} 
  = \frac{V}{4\pi^2 \alpha } \Mk^2 (c_1 + c_2) \sinh \frac{ \pi \tB}{2}
\end{equation}
where $V$ is the volume of the three dimensional space.  

The energy of the configuration is given by 
\begin{equation}
  \frac{\kappa}{2} \int\! d^3x\,dz\,  \(\frac{1}{2} h(z) B^2 
  + \Mk^2 k(z) \( F_{3z}F^{3z} - \hF_{z0} \hF^{z0}\)\).
\end{equation}
The piece proportional to $B$ gives an infinite contribution.  
This is the expected divergent energy of a space-filling magnetic field.  
This piece is independent of the constants $c_1$ and $c_2$, 
so does not affect our minimization problem. 

Performing the integrals involved in the energy 
we write
\begin{equation}
  H - \mu N_B = V \Mk^2 \sinh\frac{ \pi \tB}{2} \left( \frac{2 \kappa}{\tB} 
  \cosh\frac{\pi \tB}{2} (c_1^2 + c_2^2) - \frac\mu{4\alpha \pi^2} 
 (c_1 + c_2) \right).
\end{equation}
It is simple to minimize this with respect to $c_1$ and $c_2$.  The solution is
\begin{equation}
  c_1=c_2=\frac{\mu B}{16\pi^2\kappa \Mk^2 \cosh \frac{\pi \tB}{2}}.
\end{equation}
Plugging into (\ref{F3z}) and (\ref{Fz0}), we can write the ground state 
solutions entirely in terms of the parameters of the problem,
\begin{equation}
  F^{3z} =\frac{27 \pi}{\lambda N_c} \frac{ \mu B}{\Mk^2} 
  \frac{ \cosh \( \frac{27 \pi}{2 \lambda} 
  \frac{B}{\Mk^2} \arctan z \) }{ \cosh \( \frac{27 \pi^2}{4 \lambda} 
  \frac{B}{\Mk^2} \) ( 1 + z^2)}
\end{equation}
and
\begin{equation}
  \hat{F}^{z0} = -\frac{27 \pi}{\lambda N_c} \frac{ \mu B}{\Mk^2} 
  \frac{ \sinh \( \frac{27 \pi}{2 \lambda} 
  \frac{B}{\Mk^2} \arctan z \) }{ \cosh \( \frac{27 \pi^2}{4 \lambda} 
  \frac{B}{\Mk^2} \) ( 1 + z^2)}.
\end{equation}

The energy can also be found in terms of fundamental quantities 
by plugging in the values of $c_1$ and $c_2$ into our earlier 
expression for $H$.  Doing so gives
\begin{equation}\label{energydensity-mu}
  \epsilon\equiv\frac EV = \frac{\mu^2 B}{4\pi^2 N_c} 
  \tanh \left(\frac{27 \pi^2}{4 \lambda} \frac{B}{\Mk^2}\right).
\end{equation}
One can express the energy density in terms of the baryon number
density $n_B$,
\begin{equation}\label{energydensity}
  \epsilon = \pi^2 N_c \frac{n_B^2}B
   \coth\left(\frac{27 \pi^2}{4 \lambda} \frac{B}{\Mk^2} \right).
\end{equation}
The asymptotics of~(\ref{energydensity}) at small $B$ is
\begin{equation}
  \epsilon = \frac{4\lambda N_c}{27}\,\frac{n_B^2 \Mk^2}{B^2}\,.
\end{equation}
and at large $B$ is 
\begin{equation}
  \epsilon = \pi^2 N_c \frac{n_B^2}B.
\end{equation}
Note, however, that in order to obtain the large $B$ asymptotics 
we must assume that $B/(\lambda \Mk^2) \gg 1$.  This is precisely 
the limit in which we can no longer trust the quadratic 
approximation to the DBI action.  Thus, in order to consider this limit, 
we must be thinking in the context of a bottom-up AdS/QCD model, where
higher order terms in $F$ are suppressed from the beginning.

\section{Interpretation of results}
\label{sec:interpretation}

We now try to interpret our results for small and large magnetic
field $B$.  At small magnetic fields, one can use chiral perturbation
theory.  The Hamiltonian describing the interaction of the $\pi^0$
field with the magnetic field and baryon chemical potential
is~\cite{Son:2007ny}
\begin{equation}
  H'\equiv  H -\mu N_B = \int\!d\x\, \left( \frac12 (\bm{\nabla}\pi^0)^2 
      - \frac\mu{4\pi^2f_\pi} \bm{B}\cdot\bm{\nabla}\pi^0\right)
\end{equation}
The minimum of $H'$ is achieved at
\begin{equation}
  \bm{\nabla}\pi^0 = \frac\mu{4\pi^2f_\pi} \bm{B}
\end{equation}
at which the energy and baryon number densities are
\begin{equation}
  \epsilon = \frac{\mu^2B^2}{32\pi^4 f_\pi^2}, \qquad
  n_B = \frac{\mu B^2}{16\pi^4 f_\pi^2} .
\end{equation}
The relationship between $\epsilon$ and $n_B$ is
\begin{equation}
  \epsilon = 8\pi^4 f_\pi^2 \frac{n_B^2}{B^2}.
\end{equation}
Now using the value for the pion decay constant found
in~\cite{Sakai:2004cn},
\begin{equation}
  f_\pi^2 = \frac {\lambda N_c}{54\pi^4}M_{\rm KK}^2,
\end{equation}
we reproduce the low-$B$ asymptotics of Eq.~(\ref{energydensity}) exactly.

We now show that at very large $B$ the thermodynamic relation between
$\epsilon$ and $n_B$, obtained in the approximation where one replaces
the DBI action by the Maxwell action,
approaches that of a free noninteracting gas.
Consider a system of free, noninteracting quarks of two flavors $u$
and $d$ in an external magnetic field coupled to isospin.  The charges of
the quarks are $1/2$ and $-1/2$.  In a magnetic field, the transverse
motion is quantized.  The lowest Landau level consist of two branches,
\begin{equation}
  E = \pm k_z,
\end{equation}
each having degeneracy in the transverse direction equal to
$|e|B/(2\pi)=B/(4\pi)$.

A baryon chemical potential $\mu$ corresponds to a quark chemical
potential $\mu_q=\mu/N_c$.  The quarks fill energy levels below
$\mu_q$, leading to a nonzero baryon densty.  At small chemical
potentials (or, equivalently, large magnetic field) the filled energy levels are all in the lowest Landau
level.  The total baryon number is then
\begin{equation}
  n_B = N_f \frac B{4\pi} 
  \int\limits_{-\mu_q}^{\mu_q}\! \frac{dk_z}{2\pi} 
      = \frac{\mu B}{2\pi^2 N_c}
\end{equation}
Inverting the relation,
\begin{equation}
  \mu = \frac{2\pi^2N_c}B n_B
\end{equation}
and integrating over $n_B$, one finds the energy density as a function
of the baryon number density
\begin{equation}
  \epsilon = \pi^2 N_c \frac{n_B^2}B.
\end{equation}
This coincides exactly with the thermodynamics of our bottom-up 
model at large $B$.  Therefore, we conclude that at large $B$, the
equation of state of matter at finite-baryon density is identical to
that of free quarks.

Moreover, by redoing the previous calculation, one can show that this
feature is independent of the choice of the function $k(z)$, given
that the integral $\int_0^\infty\!dk\, k^{-1}(z)$ is convergent.  
This condition is satified in the model of~\cite{Son:2003et} 
where $k(z)\sim\cosh(bz)$. 
Therefore, in bottom-up holographic models of QCD where the action
contains only the Maxwell and Chern-Simons parts, and where chiral
symmetry breaking is due to a boundary condition at $z=0$, the
equation of state at very high magnetic field approaches that of a
free gas.

\section{Conclusions}
\label{sec:conclusion}

To conclude, we have found a solution to the field equations in the
Sakai-Sugimoto model that corresponds to matter at finite baryon
density in an external magnetic field.  In contrast to the case
without a magnetic field, it is possible to write down a solution
that is completely homogeneous in space.  At small $B$ and small
chemical potential the result can be obtained from the chiral
Lagrangian with the Chern-Simons term.  The solution continues to exist
for any value of $B$, although for larger values of $B$ the solution 
can no longer be viewed as arising from the full Sakai-Sugimoto model.  
We can, however, interpret our solution at large $B$ as arising 
from a bottom-up AdS/QCD model.  At large $B$ the system behaves, 
from the point of view of zero-temperature thermodynamics, as a system of free
quarks.

How one can explain the latter fact?  Right now we have only some
vague ideas of how it can be understood.  In magnetic fields the
fermions move in Larmor orbits whose radius shrinks as $B\to\infty$.
This fact may explain why interaction between quarks do not seem to
play any role at large $B$.

Clearly, the situation should be investigated further.  One question
one can ask is whether the state is stable with respect to small
perturbations.  At small $\mu$, the energy per baryon is small and the
system is clearly more stable than the ordinary Skyrmion crystal.  We
leave the investigation of the stability of the system at large $B$
and $\mu$ to future work.

We thank O.~Bergman for discussions.  EGT is supported, in part, by
DOE Grant No.\ DE-FG02-96ER40956.  DTS is supported, in part, by DOE
Grant No.\ DE-FG02-00ER41132.


\begin{thebibliography}{99}

\bibitem{Maldacena:1997re}
  J.~Maldacena,
  ``The large N limit of superconformal field theories and supergravity,''
  Adv.\ Theor.\ Math.\ Phys.\  {\bf 2}, 231 (1998)
  [Int.\ J.\ Theor.\ Phys.\  {\bf 38}, 1113 (1999)]
  [arXiv:hep-th/9711200].

\bibitem{Gubser:1998bc}
  S.~S.~Gubser, I.~R.~Klebanov and A.~M.~Polyakov,
  ``Gauge theory correlators from non-critical string theory,''
  Phys.\ Lett.\  B {\bf 428}, 105 (1998)
  [arXiv:hep-th/9802109].

\bibitem{Witten:1998qj}
  E.~Witten,
  ``Anti-de Sitter space and holography,''
  Adv.\ Theor.\ Math.\ Phys.\  {\bf 2}, 253 (1998)
  [arXiv:hep-th/9802150].

\bibitem{Sakai:2004cn}
  T.~Sakai and S.~Sugimoto,
  ``Low energy hadron physics in holographic QCD,''
  Prog.\ Theor.\ Phys.\  {\bf 113}, 843 (2005)
  [arXiv:hep-th/0412141].

\bibitem{Kim:2007zm}
  K.~Y.~Kim, S.~J.~Sin and I.~Zahed,
  ``The chiral model of Sakai-Sugimoto at finite baryon density,''
  JHEP {\bf 0801}, 002 (2008)
  [arXiv:0708.1469 [hep-th]];
  ``Dense holographic QCD in the Wigner-Seitz approximation,''
  arXiv:0712.1582 [hep-th].

\bibitem{Rozali:2007rx}
  M.~Rozali, H.~H.~Shieh, M.~Van Raamsdonk, and J.~Wu,
  ``Cold nuclear matter in holographic QCD,''
  J.\ High Energy Phys.\ 01 (2008) 053.
  [arXiv:0708.1322 [hep-th]].

\bibitem{Son:2007ny}
  D.~T.~Son and M.~A.~Stephanov,
  ``Axial anomaly and magnetism of nuclear and quark matter,''
  Phys.\ Rev.\  D {\bf 77}, 014021 (2008)
  [arXiv:0710.1084 [hep-ph]].

\bibitem{Bergman:2008sg} 
  O.~Bergman, G.~Lifschytz and M.~Lippert, 
  ``Response of holographic QCD to electric and magnetic fields,'' 
  arXiv:0802.3720 [hep-th]. 

\bibitem{Johnson:2008vn} 
  C.~V.~Johnson and A.~Kundu, 
  ``External fields and chiral symmetry breaking in the 
  Sakai-Sugimoto Model,'' 
  arXiv:0803.0038 [hep-th]. 

\bibitem{Son:2003et}
  D.~T.~Son and M.~A.~Stephanov,
  Phys.\ Rev.\  D {\bf 69}, 065020 (2004)
  [arXiv:hep-ph/0304182].

\bibitem{Erlich:2005qh}
  J.~Erlich, E.~Katz, D.~T.~Son and M.~A.~Stephanov,
  ``QCD and a holographic model of hadrons,''
  Phys.\ Rev.\ Lett.\  {\bf 95}, 261602 (2005)
  [arXiv:hep-ph/0501128].

\bibitem{Da Rold:2005zs}
  L.~Da Rold and A.~Pomarol,
  ``Chiral symmetry breaking from five dimensional spaces,''
  Nucl.\ Phys.\  B {\bf 721}, 79 (2005)
  [arXiv:hep-ph/0501218].

\bibitem{Bergman}
  O.~Bergman, G.~Lifschytz, and M.~Lippert, 
  arXiv:0806.0366.

\bibitem{Witten:1998zw}
  E.~Witten,
  ``Anti-de Sitter space, thermal phase transition, and confinement in 
  gauge theories,''
  Adv.\ Theor.\ Math.\ Phys.\  {\bf 2}, 505 (1998)
  [arXiv:hep-th/9803131].

\bibitem{Karch:2002sh}
  A.~Karch and E.~Katz,
  ``Adding flavor to AdS/CFT,''
  JHEP {\bf 0206}, 043 (2002)
  [arXiv:hep-th/0205236].

\bibitem{Sakai:2005yt}
  T.~Sakai and S.~Sugimoto,
  ``More on a holographic dual of QCD,''
  Prog.\ Theor.\ Phys.\  {\bf 114}, 1083 (2006)
  [arXiv:hep-th/0507073].

\bibitem{Hata:2007mb}
  H.~Hata, T.~Sakai, S.~Sugimoto and S.~Yamato,
  ``Baryons from instantons in holographic QCD,''
  arXiv:hep-th/0701280.



\end{thebibliography}
\end{document}